\documentclass[%
 aip,
 amsmath,amssymb,
 reprint,%
]{revtex4-1}
\usepackage{xcolor}
\definecolor{Black}{rgb}{0,0,0}

\usepackage{graphicx}
\usepackage{dcolumn}
\usepackage{bm}
 \usepackage[version=4]{mhchem}

\usepackage[utf8]{inputenc}
\usepackage[T1]{fontenc}
\usepackage{mathptmx}
\usepackage{etoolbox}

\makeatletter
\def\@email#1#2{%
 \endgroup
 \patchcmd{\titleblock@produce}
  {\frontmatter@RRAPformat}
  {\frontmatter@RRAPformat{\produce@RRAP{*#1\href{mailto:#2}{#2}}}\frontmatter@RRAPformat}
  {}{}
}%
\makeatother
\begin{document}

\preprint{AIP/123-QED}

\title[]{Characterisation of ferroelectric domains in magnetite (Fe$_3$O$_4$)}
\author{S. D. Seddon}
 \affiliation{
 University of Warwick, Department of Physics, Coventry, CV4 7AL
 }
 \affiliation{TU Dresden, Institute of Applied Physics, Nöthnitzer Straße 61, 01187 Dresden, Germany}
\email{samuel.seddon@tu-dresden.de}

\author{A. Cooper}%
 \affiliation{
University of Warwick, Department of Physics, Coventry, CV4 7AL
}

\author{T. Fricke}
\affiliation{Martin-Luther-Universität Halle-Wittenberg, Institut für Chemie, Anorganische Chemie
Kurt-Mothes-Straße 2, D-06120 Halle}

\author{S. G. Ebbinghaus}
\affiliation{Martin-Luther-Universität Halle-Wittenberg, Institut für Chemie, Anorganische Chemie
Kurt-Mothes-Straße 2, D-06120 Halle}

\author{M. Walker}%
 \affiliation{
University of Warwick, Department of Physics, Coventry, CV4 7AL
}

\author{T. P. A. Hase}
 \affiliation{
University of Warwick, Department of Physics, Coventry, CV4 7AL
}

\author{W. J. A. Blackmore}%
 \affiliation{
Department of Chemistry, School of Natural Sciences,The University of Manchester, Oxford Road, Manchester, M13 9PL, UK
}

\author{M. Alexe}
 \affiliation{
University of Warwick, Department of Physics, Coventry, CV4 7AL
}
\email{m.alexe@warwick.ac.uk}
\date{\today}

\begin{abstract}
Magnetite has long been investigated across many disciplines due to the interplay between its ferroic order parameters, namely its ferrimagnetism, ferroelasticity and ferroelectricty. Despite this, the experimental difficulty in measuring low temperature real space images of the ferroelectric domains has meant that the local behaviour of ferroelectric domains emergent below the $\sim 38~\text{K}$ phase transition have yet to be realised. This work presents real space images of the ferroelectric domains, and uses piezo force microscopy to as a function of temperature to probe the onset of piezoelectricty and ferroelectricity across the 38~K transition. 
\end{abstract}

\maketitle

Magnetite, Fe$_3$O$_4$, is one of mankinds oldest functional materials, going through its first ferroic ordering temperature at 860~K where it becomes ferrimagnetic  \cite{alexe_ferroelectric_2009, ziese_magnetite_2012}. It is of distinct interest to the paleomagnetic community, principly whilst studying the magnetisation of the ocean floor. Magnetic stripes emerging in magnetite on the Earth's crust caused by the Earth's magnetic pole reversal led to a resurgence and eventual acceptance of the theory of continental drift \cite{vine_magnetic_1963}. Beyond terrestial applications, magnetite has also been shown to possibly hold the secrets to life on Mars, due to its presence in martian meteorites \cite{mckay_search_1996}.

Magnetite also has significant spintronics applications due to its low temperature electronic phase transition, at 120~K, identified by Verwey in 1941 \cite{verwey_electronic_1941}. Across this Verwey transition there is not only a large change in the electrical resistivity and magnetic moment of magnetite, but there is also the onset of ferroelasticity. There is a corresponding change in crystal structure \cite{chen_microstructure_2008,wright_charge_2002,rozenberg_origin_2006} as the crystal structure changes from an inverse spinnel/ cubic $Fd\overline{3}m$ symmetry to the monoclinic, \textit{Cc} space group \cite{novak_nmr_2000,zuo_charge_1990,joly_low-temperature_2008,angst_intrinsic_2019,de_la_figuera_magnetite_2019}. The exact nature of this phase transition has been hotly debated, however the charge ordering was finally described in 2012 by \citet{senn_charge_2012}.

At 38 K there is a further phase transition with the introduction of ferroelectricity. Initially shown with macroscopic DC applied field hysteresis loops \cite{rado_electric_1975,kato_observation_1982, miyamoto_crystal_1988,miyamoto_ferroelectricity_1994,kato_ferrimagnetic_1983}, it was later verified by \citet{alexe_ferroelectric_2009} who used a positive up negative down (PUND) measurement regime to determine the onset of polarisation as a function of temperature, on thin film \ce{Fe3O4}. Following this, \citet{ziese_magnetite_2012} showed that the behaviour of magnetite below 38~K was similar to that of a relaxor ferroelectric, by exploring the frequency dependence of electrical permittivity in thin films. Identical behaviour was also found in single crystals \cite{kobayashi_dielectric_1986, schrettle_relaxor_2011}, where a frequency dependence of the permittivity equally confirmed relaxor behaviour. 

The crystal symmetry in the low temperature phase ($<$38~K) is triclinic \cite{medrano_domains_1999,miyamoto_crystal_1988}. This non-centrosymmetric symmetry fulfils the structure requirements for the measured ferroelectricity mechanisms under pinning the ferroelectricity, which have been ascribed to small alterations in charge density and bond lengths caused by further charge ordering \cite{ziese_magnetite_2012,brink_multiferroicity_2008,yamauchi_ferroelectricity_2009}.

This work uses low temperature piezo force microscopy (PFM) on a single crystal of Fe$_3$O$_4$(001) to explore the nature of the ferroelectricity around and below the ferroelectric phase transition at 38 K. The first real space images of the ferroelectric domains are presented, as well as local investigations of the ferroelectric phase transition, and explorations of potential multiferroic couplings.

\begin{figure*}
\includegraphics[width = \textwidth]{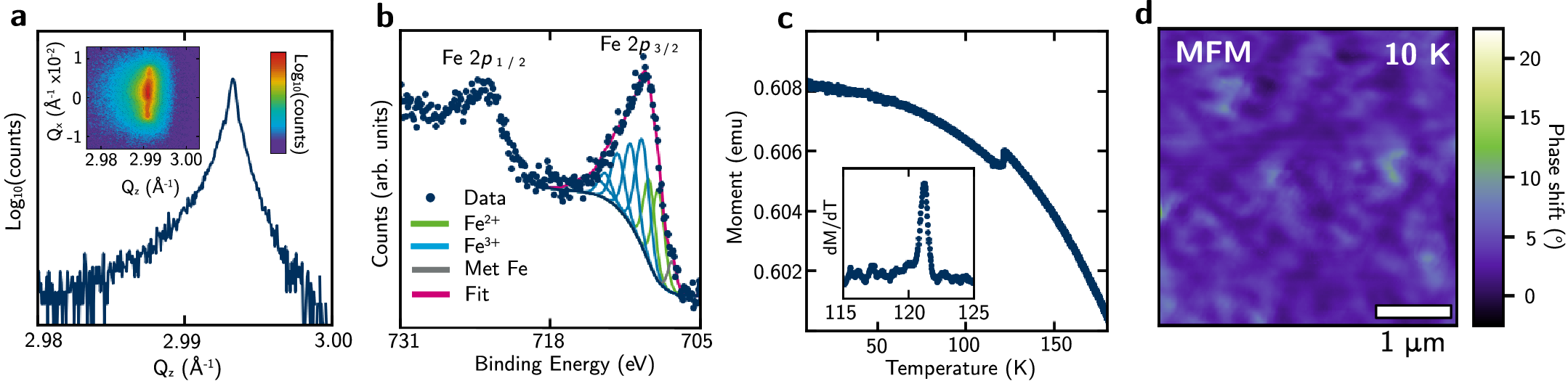}
\caption{\label{fig:Fe3O4_samp_char}\textbf{a} X-ray diffraction (High resolution 2$\theta$ - $\omega$ and reciprocal space map (RSM) (Inset)) acquired of the surface normal 004 peak, verifying crystallinity of the sample.  \textbf{b} X-ray photoelectron spectroscopy (XPS) with the Fe 2\textit{p}$_{3/2}$ fitted in accordance with the literature confirming sample stoichiometry. \textbf{c} Vibrating Sample Magnetometry (VSM) data of magnetisation versus temperature confirming the presence of a Verwey transition. \textbf{(inset)} differential of the $M(T)$ detailing the Verwey transition to be 121 K as expected. \textbf{d} Magnetic Force Microcsopy image acquired at 10 K.}
\end{figure*}

Magnetite single crystals were grown by the optical floating zone method using Fe$_2$O$_3$ (Acros Organics, 99.999~\% purity) as the starting material. Further details can be found in the suplemntary information.
Piezo force microscopy (PFM) images were acquired on a low temperature AFM (Attocube attoLiquid 2000) provided with interferrometric detection system.
Nanosenors PPP-EFM tips, ~80 KHz resonance, were used.

In order to verify the quality of the crystal we performed a series of sample characterisation measurements. Figure~\ref{fig:Fe3O4_samp_char}\textbf{a} shows a surface normal X-ray trunction rod of the 004 specular peak, indicating a surface normal along the 001 cubic direction. The peak position is at $Q_z=2.995 $\AA$ ^{-1}$ which corresponds to a \textit{d} spacing of $2.098\  $\AA\  as expected at room temperature. Some asymmetry can be seen in the peak shape,
likely introduced from the repeated thermal cycling from magnetometry measurements completed in advance. A reciprocal space map (RSM) (in the inset of Figure~\ref{fig:Fe3O4_samp_char}(a)) of the same peak was acquired, where this asymmetry is more apparent.  A considerable broadening in $Q_x$ is seen relative to $Q_z$, a mosaic broadening typical of polished samples.  Given that this broadening occurs along $Q_x$, i.e. the in plane direction, it does not itself affect the determination of the out of plane normal.
In both cases, the peak corresponds to a crystal a-axis of $8.392\  $\AA, which is a strong indicator of the purity and crystallininity of the sample (the reported lattice parameter of magnetite is $8.3893\ $\AA\  \cite{Fonseca_2016}). 

Identifying the oxidation state of iron is key to confirming the stoichiometry, especially due to the high sensitivity to oxygen whilst growing magnetite \cite{gilks_polar_2016,zhang_oxygen_2017}. Other stoichiometries frequently appear as parasitic phases, notably Fe$_2$O$_3$ and FeO. 
 As seen in figure \ref{fig:Fe3O4_samp_char}b, the highly constrained component peaks fit well\cite{biesinger_resolving_2011} to Fe 2\textit{p}$_{3/2}$ data indicating that the sample is Fe$_3$O$_4$ with no apparent parasitic phases. The presence of a metallic Fe component is most likely caused by the presence of surface defects. Such defects are unavoidable and common even in mass produced commercially available Si or GaAs wafers \cite{seddon_work_2019}.

Identification of the Verwey transition by $M(T)$ or resistivity measurements is a common approach to confirm sample purity. A very low remenance and magnetic coersivity was found when probing the magnetic field dependence. To check the expected magnetic properties in the sample, magnetisation as a function of temperature ($M(T)$) measurements were conducted using a Quantum Design MPMS3 SQUID magnetometer, seen in figure \ref{fig:Fe3O4_samp_char}c. A magnetic field of 3 T was applied along the axial surface normal ([001]) and the magnetisation was measured as a function of temperature, starting at 200 K down to 2 K. As can be seen in Figure~\ref{fig:Fe3O4_samp_char}(c) a sharp drop in the magnetisation is seen at around 120 K, consistent with the Verwey transition, even under field cooling. The inset displays the differential of magnetisation with respect to temperature, clearly showing a sharp spike at 121 K. 

No magnetic field dependence on ferroelectricity was found later in the study, however to further confirm the magnetic domain behaviour on the nanoscale, magnetic force microscopy (MFM) data were recorded and presented in Figure~\ref{fig:Fe3O4_samp_char}(d). Expected phase shift values for magnetic domains in this material were established across various temperatures, and are the subject of a later work. However, in the data at 10~K, very low phase shifts were seen, indicating magnetic domains lie almost entirely in the same direction, which is likely in-plane. Application of a magnetic field revealed magnetic domain switching on a hierarchy far above the ferroelectic domains reported below, and no correlation between the two was eventually found. 

\begin{figure}
    \includegraphics[width = 0.49\textwidth]{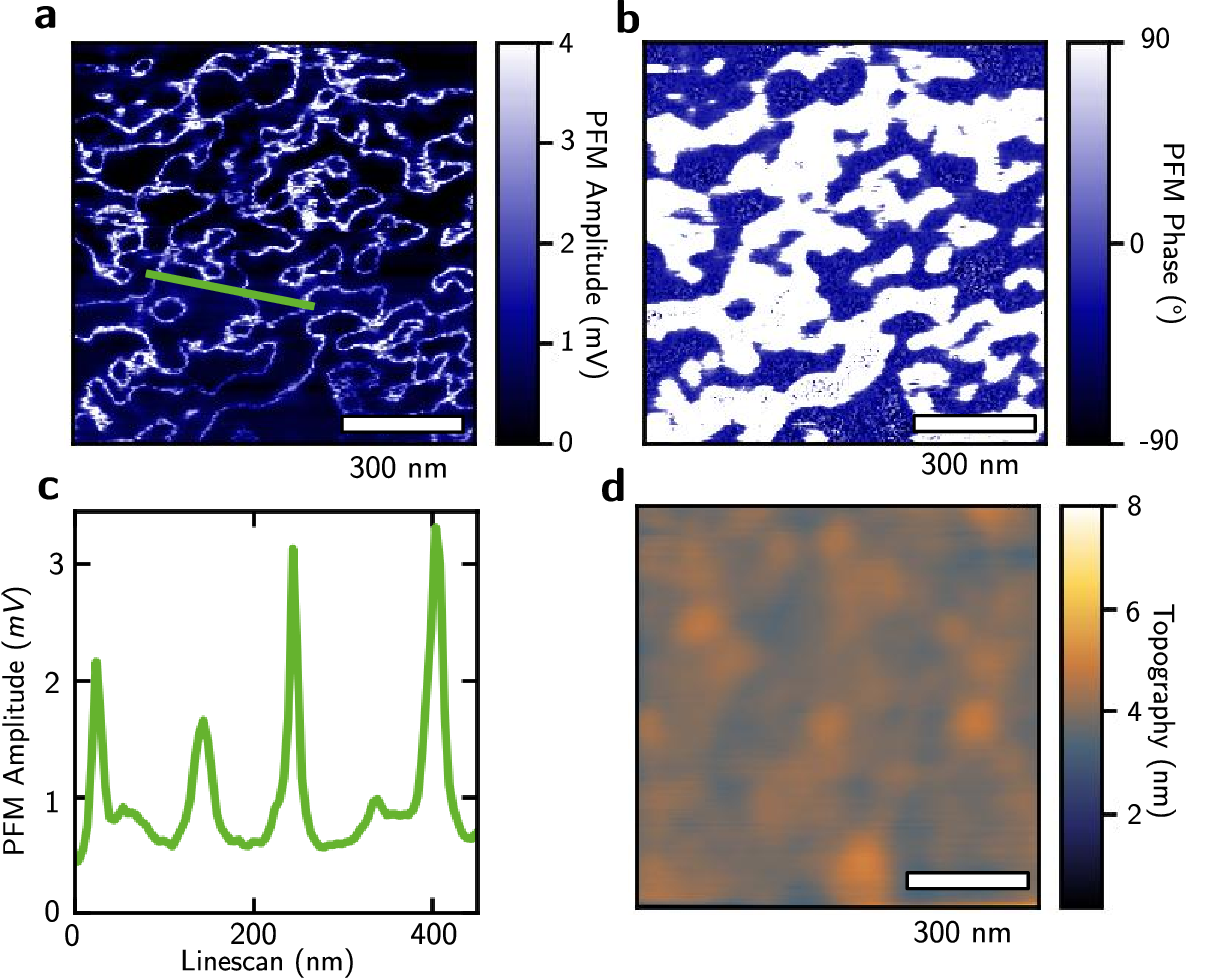}
    \caption{\label{fig:PFM_images}\textbf{a} and \textbf{b} PFM amplitude and phase measurements acquired at 10~K. \textbf{c} a line scan as marked in a. \textbf{d} corresponding topography measurement simultaneously acquired as in\textbf{a} and \textbf{b}.}
    \label{fig:Fe3O4-PFM}
\end{figure}

Despite ferroelectricty already being confirmed in magnetite samples from bulk measurements, there currently exists no local measurements of ferroelectric domain patterns or local confirmations of ferroelectricity in the literature. Futhermore, due to the complexity of the crystal symmetry (triclinic), any new real space images of how ferroelectric domains manifest in these systems would be a step towards begining to understand the complex expression of ferroelectricty in such low symmetry systems.  Piezo force microscopy (PFM) is a well established surface analysis technique that measures not only the amplitude of oscillating piezoelectric domains but also the phase of the oscillating domains. Shifts in phase are indicative of different underlying ferroelectric domains. Here, measurements acquired at 10~K are presented in Figure~\ref{fig:Fe3O4-PFM}. In order to preserve the quality of the tip prior to scanning a flat region was identified with non-contact topographic AFM measurements. 

Due to the low amplitude signal of the ferroelectric response from the magnetite surface, the PFM signal was recorded in resonance mode.\cite{Harnega2003} The amplitude signal acquired can be seen in Figure \ref{fig:Fe3O4-PFM}a, where contrary to the traditional expression of ferroelectric domains imaged by PFM measurements quite the opposite is seen. The initial impression is a distribution of narrow, randomly distributed and somewhat circular domains. Differences in individual domains is most apparent in the PFM phase images in figure~\ref{fig:Fe3O4-PFM}\textbf{b} where two distinct PFM phases are identified, approximately 180$^o$ apart. The PFM amplitude on the domains themselves is low, but relatively constant, the domain walls are highlighted by an increased PFM amplitude response which can be observed in the line scan in figure~\ref{fig:Fe3O4-PFM}\textbf{c}.

This behaviour in the PFM amplitude in magnetite is atypical of traditional measurements. Instead of a constant amplitude on the domains, with a dip in amplitude as the tip crosses the domain wall, here amplitude signal is constant, with the occasional spikes outlining small, closed domains. These spikes in amplitude is highlighted by the linescan in \ref{fig:Fe3O4-PFM}c where the enhancement can be seen as 5-6 times larger than the normal response. 

Interpretation of the amplitude signals alone may indicate that the ferroelectricity in the material is somehow constrained to the domain walls. However, analysis of the phase images indicates otherwise. The phase image of the PFM measurements describes a more traditional behaviour as can be seen in Figure~\ref{fig:Fe3O4-PFM}b. Two distinct domains are resolved, corresponding to the regions outline by amplitude measurements. The PFM phase image, combined with amplitude measurements indicate that different domains present on the magnetite surface have a similar, indistinguishable amplitude but different phase, implying that the ferroelectric polarisations consistently possess an out of plane component in different directions, switching the phase. The small size of the domains relative to most ferroelectrics is consistent with the relaxor nature of magnetite\cite{miao_ferroelectricity_2016, Belhadi2021}

The spiking in amplitude at domains walls needs careful treatment as there is some evidence it could arise from a scanning artefact. For example, there have been reports that the positioning of the detection laser on the back of the cantilever can directly affect the contrast seen on ferroelectric domains. Indeed there is one position on the cantilever that causes a similar domain wall response, spiking as opposed to dropping to nearly zero amplitude \cite{kim_electrostatic-free_2017}. 

Another cause for enhanced amplitude could be a mechanical softening of the material occurring at the domain walls. This is a phenomenon is well documented in ferroelectric materials by \citet{stefani_mechanical_2020}, with the mechanical softening causing a shift in the resonant frequency of the tip-sample system \cite{stefani_mechanical_2021}. When resonating at a fixed frequency, a shift in the resonance of the system could result in a sharp change in the amplitude over the region softened.  Contrary arguments to this are that if the shift of the resonance causes a spike in recorded amplitude, in principle choosing sequential measurement frequencies through the resonance should cause the opposite effect, with enhanced amplitude occurring at the domains, and a reduced amplitude signal occurring at the domain walls. Despite this, the effect was shown to be frequency independent. Similarly dual resonance frequency tracked PFM (DFRT) was also employed in order to attempt to minimise this spiking of amplitude to no success, with images showing no difference in contrast. As such, the amplitude spikes likely not caused interctions with the topography (an effect DFRT is designed to mitigate) but instead arise due to a real effect of the system itself. 

Seeminly not a measurement artefact, we attribute the enhanced amplitudes at domain walls seen as instead be a real effect, caused by a shear movement of the ferroelectric domains at the domain wall. Unlike (near) uniaxial ferroelectics such as LiNiO$_3$ \cite{godau_enhancing_2017,zhao_high_2020} or \ce{BiFeO3} \cite{folkman_stripe_2009,nelson_spontaneous_2011}, the crystal symmetry of Fe$_3$O$_4$ is triclinic. When considering a cubic system, the piezoelectric tensor is diagonal. As crystal symmetry is reduced, the number of non-zero terms in this tensor increase. Due to the complexity of the unknown piezoelectric tensor, correlation with a direct piezoelectric response is extremely challenging.

\begin{figure}
    \includegraphics[width = 0.49\textwidth]{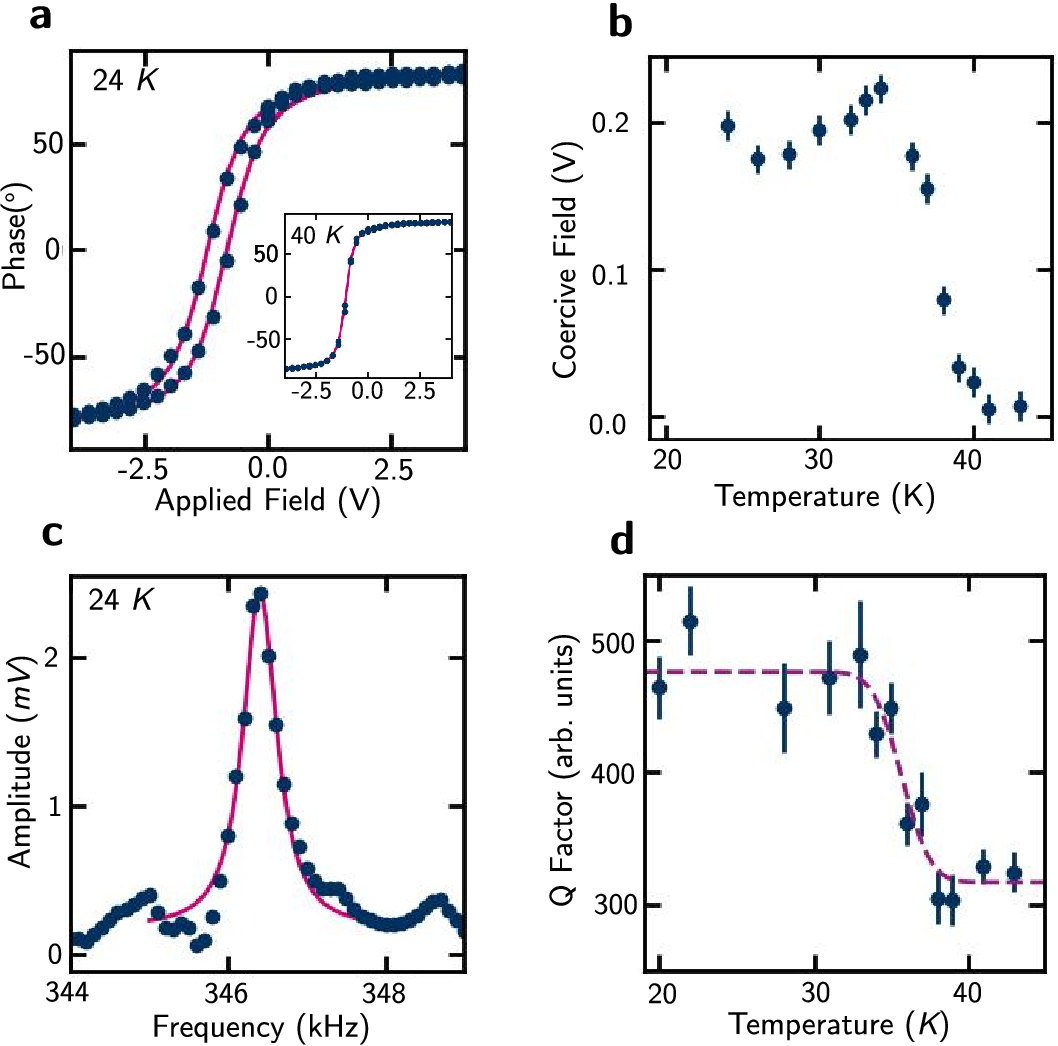}
    \caption{\textbf{a} Typical local hysteresis in the cantilever phase acquired by applying ±4 V showing an open loop. (\textbf{Inset}) example of a closed loop acquired at 40 K, above the ferroelectric phase transition. \textbf{b} Temperature dependence of the electric coersive field measured by AFM tip, from fitting hysteresis loops as in \textbf{a}. \textbf{c} Example of a typical contact resonance acquired at 24 K, fitted with a Pearson VII peak function. \textbf{d} Temperature dependence of Q-factor of contact resonance interaction with the surface. Line is a guide to the eye.}
    \label{fig:Fe3O4-PhaseT}
\end{figure}

Whilst ferroelectricty across the 38 K phase transition has been established macroscopically, so far this was not established locally. The most typical example of local determination of ferroelectricity is to measure the effect on PFM amplitude and phase when DC electric field is applied to the sample, creating a ferroelectric hystersis loop. The result of applied electric field on the cantilever phase is presented in Figure \ref{fig:Fe3O4-PhaseT}\textbf{a}
where a hysteresis loop was applied first up to 4 V, then down through 0~V to -4 V before finally returning to 0 V. 

The main loop in Figure \ref{fig:Fe3O4-PhaseT}\textbf{a} acquired at 24 K, well below the ferroelectric phase transition show a clear opening in the loop, indicating the presence of a coersive electric field arising from the switching of a ferroelectric polarisation in the material. Contrary to this, as can be seen in the 40 K inset, above the phase transition a typical paraelectric closed hysteresis loop where there exists. The centre of the hysteresis is shifted by approximately -1.5 V from the centre of the applied field/ Such an effect is not uncommon, with the direct interpretation being surface induced anisotropy, implying it would rather lie in one particular direction. This kind of imprinting has been seen before\cite{alexe_polarization_2001}. Data were fitted with a modified Langevin function, commonly used to fit magnetic hysteresis data (Figure~\ref{fig:Fe3O4-PhaseT} \textbf{a}) \cite{procter_magnetic_2015,thorarinsdottir_amorphous_2021} of the form 
\begin{equation}
    \phi(E) = \phi_{\text{max}} \left( \frac{1}{\tanh(\frac{E+E_c}{S})} - \frac{1}{\frac{E+E_c}{S}} \right),
\end{equation}
where $\phi(E)$ is the phase as a function of applied electric field, $\phi_{\text{max}}$ is the saturation value of $\phi(E)$, $E$ is the applied field, $E_c$ is the coercive field and $S$ is a shape parameter of the loop, from which no physical properties will be determined. The temperature dependence of $E_c$ is presented in Figure~\ref{fig:Fe3O4-PhaseT}\textbf{b}. Above the bulk-identified phase transition temperature the coersive field of the system is zero within error bars (determined from sensitivity of the measurement apparatus), however as the temperature is reduced past 38 K, a rise in the coersive field is sharp but continuous as expected of a second order phase transition, and consistent with previous measurements \cite{alexe_ferroelectric_2009}. This coersive field increases sharply with decreasing temperature, before settling to a relatively stable value of around 0.2~V. Due to the fiberoptic attocube system being sensitive to the out of plane tip vibrations only, the measurement is only sensitive to the out-of-plane piezoelectric oscillations (effectively the d$_{zz}$ or out-of-plane piezoelectric coefficient). Lack of in-plane resolution prevents further quantification of the polarisation on a local scale however the general behaviour is in agreement with macroscopic measurements. 

Modification of the elastic properties of the sample surface (such as the onest of piezoelectricity) shall induce a change in the tip-surface resonance. As shown by \citet{Salje:pd5021}, most ferroelastic materials soften as the system undergo a structural phase transition. Therefore, in the case of magnetite undergoing the 38~K ferroelectric transition, the Q-factor of the tip sample resonance is expected to increase as the temperature decreases across the phase transition region. An example of a typical contact resonance can be found in Figure~\ref{fig:Fe3O4-PhaseT}\textbf{c}, where data has been fitted with a Pearson VII function to determine exact parameters such as peak width and position. 
Whilst small features either side of the central peak greatly affect the fit at the fringes of the peak function, the model clearly captures well the upper parts of the central resonance peak $f_r$, as well as the full width half maximum, $w$. With these two parameters a Q factor can determined as $\frac{f_r}{w}$ and allow the interaction to be compared as a function of temperature. 

The temperature dependence of the Q-factor is seen in Figure~\ref{fig:Fe3O4-PhaseT}\textbf{d}, with individual Q factors and their errors determined from fitting. As in the coersive field data, an initially stable Q factor is shown to stiffen as the temperature of the system is reduced past the 38 K ferroelectric phase transition, before reaching a stable value below 34~K, suggesting not just a structural phase transition but also the onset of piezoelectricity.  

The ferroelectric properties of magnetite \ce{Fe3O4} were probed across its ferroelectric phase transition at 38 K. The first PFM images of the ferroelectric domains reveal unusual spiking behaviour of the PFM amplitude at domain walls. Different domains themselves are revealed by the PFM phase recorded however the amplitude at these domains themselves is relatively low. The unusual behaviour is ascribed to shear domain wall motion caused by the triclinic crystal symmetry. The opening of locally acquired hysterisis loops was probed locally by PFM and shown to open at 38~K, settling to a value of approximately 0.2~V coersive field below 30 K. A stiffening between the tip-sample system was also found after observation of the contact resonance of the tip, indicating the onset of piezoelectricty upon the cooling of the system through the ferroelectric phase transition. 

\textbf{Data Availability Statement}
The data that support the findings of this study are available from the corresponding author upon reasonable request.

\textbf{Conflicts of Interest}
The authors declare no conflicts of interest. 

\textbf{Acknowledgements}
WJAB thanks the European Research Council (ERC-2019-STG-851504) - PI: Nicholas. F Chilton, The University of Manchester, for funding. 
MW acknowledges financial support from the EPSRC-funded Warwick Analytical Science Centre (EP/V007688/1). We acknowledge the EPSRC UK National Electron Paramagnetic Resonance Service for access to the SQuID magnetometer (EP/S033181/1). We thank Adam Brookfield and Dr Jordan Thompson for technical assistance. The financial support by EPSRC through grants  EP/M022706/1 and EP/T027207/1 is gratefully acknowledged. M.A. acknowledges also the Alexander von Humboldt Research Award.

\bibliography{fer_mag.bib}

\end{document}